\documentclass[prb,twocolumn,superscriptaddress,floatfix,showpacs]{revtex4}
\pdfoutput=1
\usepackage{graphicx,amsfonts,amssymb,amsmath, hyperref}

\newif\ifhyper
\hypertrue
\ifhyper
\hypersetup{
   citecolor = {green},
   colorlinks = {true}, % false
   urlcolor = {blue} % magenta
}
\fi

\newcommand{\beq}{\begin{equation}}
\newcommand{\eeq}{\end{equation}}
\newcommand{\beqa}{\begin{eqnarray}}
\newcommand{\eeqa}{\end{eqnarray}}
\newcommand{\ket} [1] {\vert #1 \rangle}

\newcommand{\braket}[2]{\langle #1 | #2 \rangle}

\def\ket#1{\vert#1\rangle}

\def\ipr#1#2{\langle#1\vert#2\rangle}

\def\Longarrow{\protect\@lra}
\def\@lra{\relbar\joinrel\relbar\joinrel\relbar\joinrel%
          \relbar\joinrel\rightarrow}

\usepackage[utf8]{inputenc}
\usepackage[T1]{fontenc}
\usepackage{lmodern}
\usepackage{microtype}
\usepackage{mathtools}
\DeclarePairedDelimiter{\abs}{\lvert}{\rvert}
\DeclarePairedDelimiter{\set}{\{}{\}}
\DeclarePairedDelimiterX{\setbuilder}[2]{\{}{\}}{#1\mathrel{\delimsize\vert} #2}
\usepackage{amssymb}

\RequirePackage{amsthm}

\theoremstyle{definition}

\usepackage{booktabs}

\usepackage{csquotes}

\begin{document}

\title{Topological Minimally Entangled States via Geometric Measure}

\author{Oliver Buerschaper}
\affiliation{Perimeter Institute for Theoretical Physics, 31 Caroline Street North, Waterloo, Ontario, N2L\,2Y5, Canada}

\author{Artur Garc\'ia-Saez}
 \affiliation{C. N. Yang Institute for Theoretical Physics, State
University of New York at Stony Brook, NY 11794-3840, USA}

\author{Rom\'an Or\'us}
\affiliation{Institute of Physics, Johannes Gutenberg University, 55099 Mainz, Germany}

\author{Tzu-Chieh Wei}
 \affiliation{C. N. Yang Institute for Theoretical Physics, State
University of New York at Stony Brook, NY 11794-3840, USA}

\begin{abstract}
Here we show how the Minimally Entangled States (MES) of a 2d system with topological order can be identified using the geometric measure of entanglement. We show this by minimizing this measure for the doubled semion, doubled Fibonacci and toric code models on a torus with non-trivial topological partitions. Our calculations are done either quasi-exactly for small system sizes, or using the tensor network approach in [R. Or\'us, T.-C. Wei, O. Buerschaper, A. Garc\'ia-Saez, arXiv:1406.0585] for large sizes. As a byproduct of our methods, we see that the minimisation of the geometric entanglement can also determine the number of Abelian quasiparticle excitations in a given model. The results in this paper provide a very efficient and accurate way of extracting the full topological information of a 2d quantum lattice model from the multipartite entanglement structure of its ground states. 

\end{abstract}

\maketitle

\section{Introduction}

Any 2d quantum spin model with intrinsic topological order has a finite ground state degeneracy on a torus \cite{topo}.
This nontrivial ground state space features a distinguished basis which is given by the eigenvectors of the modular $T$-transformation ({Dehn} twist).
It is also this basis in which the topological $S$-matrix is expressed.
From the perspective of topological quantum field theory (\textsc{TQFT}) these distinguished basis states are in one-to-one correspondence with the (irreducible) charges that label the boundaries of the 2d surfaces in (2+1)d \textsc{TQFT}s.
In the corresponding lattice model, the charges label quasiparticle excitations which can be moved by string-like operators.
These operators have a surprisingly rich algebraic structure which encodes the mutual and self-statistics of the quasiparticle excitations, for example.
When wrapped around non-contractible loops, the string-like operators are also known as {Wilson} loop operators.

Furthermore, intrinsic topological order has been linked to different patterns of long-range entanglement under local unitary (LU) equivalence \cite{Chen:2010gb,Bachmann:2011kw}. For example, this manifests itself in the topological entanglement entropy which is a nonlocal quantity tied to bipartitions of ground states \cite{Hamma:2005p63,Kitaev:2006,Levin:2006ij}. Other entanglement measures show also topological contributions \cite{topoOther, topoGE1, topoGE2}. However, we do not know of a single measure which can even begin to capture all the intricate patterns of entanglement in a quantum many-body state\cite{Osterloh:2012gg}.  In other words, our understanding of multipartite entanglement in quantum states is relatively poor at present.

Suppose we have access to a complete set of linearly independent ground states for a given 2d lattice model.
Tasked with identifying its universality class we will typically face several difficulties.
Firstly, the string-like operators encoding the properties of quasiparticle excitations are not known a priori.
Secondly, it may be impossible to actually perform {Dehn} surgery on many lattices.
However, it has been argued convincingly\cite{Dong:2008ce,Zhang:2012jc,Cincio:2013ku,Zhu:2013wt,Zhu:2014jf} that the distinguished basis states~$\set{\ket{\Xi_i}}$ are singled out by their entanglement properties. Namely, they minimise the entanglement entropy of certain non-contractible regions on the torus, which is why they have been called \emph{Minimally Entangled States} (MES). Thus we may neither need to know the string-like operators nor perform explicitly {Dehn} surgery to identify a topological universality class, provided the MES can be found.

The entanglement entropy is a measure of entanglement between
a region $A$ and its complement, denoted by $A^\perp$. It is defined by
the von Neumann entropy of the subsystem $A$, i.e., $S(\rho_A) = -{\rm tr} (\rho_A \log_2 \rho_A)$. Here the reduced density matrix $\rho_A={\rm
Tr}_{A^\perp}(|\Psi\rangle\langle\Psi|)$ is obtained by tracing out 
the degrees of freedom in $A^\perp$. The entanglement entropy is bipartite in nature,
but each region $A$ or $A^\perp$ can contain many lattice sites. Rather than a bipartite measure of entanglement, here we
consider a multipartite measure, the
geometric entanglement (GE), and investigate how it can be used to
characterise a topological phase. Considering such a multipartite measure brings a number of advantages over the  ``more usual'' bipartite entanglement, especially regarding its numerical evaluation. In particular, it was shown in Ref.\cite{topoGE2} that this quantity can be computed more efficiently \emph{and} more accurately than bipartite entanglement measures by means of a tensor network algorithm. 

Motivated by the above, here we show that the GE evaluated on topologically
nontrivial partitions can be used to identify the topologically
distinguished basis of MES. This result has important implications both practical and fundamental. As said before, the
\textsc{GE} provides an alternative figure of merit which may prove
more convenient to evaluate numerically than other entanglement measures \cite{topoGE2}. Moreover, and from a fundamental perspective, the fact that the distinguished basis minimises the \textsc{GE} is a highly nontrivial statement about the structure of many-body entanglement in topologically ordered states. 

The organisation of this paper is as follows: in Sec.II we review some preliminary concepts on MES, the GE, and the models that we will study. In Sec.III we provide our results for small-size systems, focusing on the doubled semion and doubled Fibonacci models. Sec. IV deals with large sizes for the toric code model using a tensor network approach. Finally, in Sec.V we provide the conclusions of our work and some perspectives for the future.

\section{Preliminary concepts}

Here we provide a brief introduction to the two main concepts to be discussed in this paper, namely: MES for topological 2d systems,  and the geometric measure of entanglement for multipartite states. 

\subsection{Minimally Entangled States}

Let us consider a 2d topological system on a torus. As is well-known, the system will display a topological degeneracy in the ground state subspace. Let us now consider the entanglement properties of the different states in the ground subspace by, say, considering the entanglement entropy $S(\rho_A)$ of a bipartition. As is widely believed, for smooth bipartitions with a contractible boundary of perimeter $L$, the entanglement entropy behaves like 
\beq
S(\rho_A) = S_0 - S_{\gamma} + O(L^{-\nu}),
\eeq
with $S_0 \propto L$ (the so-called \emph{area-law} behaviour) and $S_{\gamma}$ the topological entanglement entropy, which for topological systems is a universal contribution and determines the presence of topological order by itself. 

For bipartitions with a non-contractible boundary, the situation is quite different \cite{Zhang:2012jc}. In such a case, the topological contribution $S_{\gamma}$ actually depends on the specific choice of ground state within the ground subspace. Those states having the maximum topological component, or equivalently the minimum overall entanglement entropy, are called \emph{Minimal Entropy States}. For the sake of this paper, since we will be dealing with other entanglement measures rather than the entropy, we shall call them \emph{Minimally Entangled States}, or MES. 

MES are interesting since, as is well-known by now \cite{Zhang:2012jc}, one can extract all the topological information about the system just from their mutual overlaps, e.g., $S$ and $T$ matrices. These states, which we call here $\set{\ket{\Xi_i}}$, are also from the distinguished topological basis discussed in the introduction. 

\subsection{Geometric Entanglement}

Consider an $m$-partite normalised pure state $\ket{\Psi} \in \mathcal{H} =
\bigotimes_{i=1}^{m} \mathcal{H}^{[i]}$, where $\mathcal{H}^{[i]}$
is the Hilbert space of party $i$. For instance, in a system of $n$
spins each party could be a single spin, so that $m = n$, but could
also be a set of spins, either contiguous (a \emph{block}
\cite{geometric2}) or not. We wish now to determine how well the state
$\ket{\Psi}$ can be approximated by an unentangled (normalised)
state of the parties,
$\ket{\Phi}\equiv\mathop{\otimes}_{i=1}^{m}|\phi^{[i]}\rangle$. The
proximity of $\ket{\Psi}$ to $\ket{\Phi}$ is captured by their
overlap. The entanglement of $\ket{\Psi}$ is thus revealed by the
maximal overlap~\cite{ge},
$\Lambda_{\max}({\Psi})\equiv\max_{\Phi}|\ipr{\Phi}{\Psi}|$. The
larger $\Lambda_{\max}$ is, the less entangled is $\ket{\Psi}$. We
quantify the entanglement of $\ket{\Psi}$ via the quantity:
\begin{equation}
E_G({\Psi})\equiv-\log_2\Lambda^2_{\max}(\Psi), \label{eq:Entrelate}
\end{equation}
where we have taken the base-2 logarithm, and which gives zero for
unentangled states. $E_G(\Psi)$ is called \emph{geometric
entanglement} (GE). This quantity has been studied in a variety of
contexts, including critical systems and quantum phase transitions
\cite{geometric2, geometric3}, quantification of entanglement as a
resource for quantum computation \cite{resource}, local state
discrimination \cite{discrim}, and has been recently measured in NMR
experiments \cite{exper}.  Also, one can choose the case of just two
sets of spins. In this case the GE $E_G(\Psi)$ coincides with the
so-called single-copy entanglement  between the two sets,
\begin{equation}
    \label{eq:single-copy}
    E_1(\Psi)
    =-\log_2
     \nu_1(\rho),
\end{equation}
with $\nu_1(\rho)$ the largest eigenvalue of the reduced density
matrix $\rho$ of either set \cite{sc}.

The GE offers a lot of flexibility to study multipartite quantum
correlations in spin systems. For instance, one can choose each
party to be a single spin, but one can also choose blocks of
increasing boundary length $L$ \cite{geometric2}. Studying how the GE changes with $L$ provides information about how
close the system is to a product state under coarse-graining
transformations. What is more, one can
choose each block to consist of spins in non-contractible regions,
which is what we shall mainly use here in the investigation of
MES.

There have been two recent notable findings regarding the GE of quantum many-body states. The first of these results is that for renormalization group (RG) fixed points such as the toric code and other topological exactly-solvable models, the GE of blocks
exactly obeys $E_G = E_0 - E_{\gamma}$, with $E_{\gamma}$ a topological
contribution (the topological GE) and $E_0$ some non-universal term
\cite{topoGE1}. Moreover, it was observed that $E_{\gamma}$ coincided with the topological contribution to the entanglement entropy for
the considered models. As for $E_0$ it was found that $E_0 \propto
n_b L$, with $n_b$ the number of blocks with a boundary
of size $L$.

The second notable result has been the development of an efficient tensor network \cite{tn} algorithm based on Projected Entangled Pair States \cite{PEPS} to compute the GE for a torus partitioned into cylinders \cite{topoGE2}. This method was used to find sharp evidence of topological phase transitions in 2d systems with a string-tension perturbation. In fact, when compared to tensor network methods for R\'enyi entropies, this approach turned out to produce almost perfect accuracies close to criticality and, on top, was orders of magnitude faster than more ``standard'' R\'enyi entropy calculations.

In what follows we will show that the GE of topological ground states, when computed for a torus partitioned into cylinders, can be used to determine the distinguished topologically-preferred basis of MES. For systems with a small size we are able to do this almost exactly, i.e., without the need of any tensor network implementation. For larger systems, however, we make use of the algorithm introduced in Ref.\cite{topoGE2}. When combined with such a tensor network approach, the overall algorithm turns out to be a very efficient and precise way of identifying the MES for a topological 2d system. 

{
Before we move on to describe the topological models, let us briefly discuss how one computes the GE in general. Here we describe an iterative
method to compute the maximal overlap, which has been described previously in the Supplemental Material of Ref.~\cite{exper} and was also discussed in Ref.~\cite{Bruss}. This
method can not only be implemented numerically, but can also be
carried out experimentally~\cite{exper}. To compute the maximal overlap for the
state $|\Psi\rangle$ with respect to product states
$|\Phi \rangle\equiv\bigotimes_{i=1}^{m}|\phi^{[i]}\rangle$, we use
the Lagrange multiplier $\lambda$ to enforce the constraint
$\langle \Phi|\Phi\rangle=1$,
\begin{equation}
f(\Phi)\equiv \langle \Phi|\Psi\rangle\langle \Psi |\Phi\rangle -\lambda
\langle \Phi|\Phi\rangle.
\end{equation}
Maximizing $f$ with respect to the local product state
$|\phi^{[i])}\rangle$, we obtain the extremal condition, originally derived in Ref.~\cite{ge},
\begin{equation}
{\cal H}_{\rm eff}^{[i]}|\phi^{[i]}\rangle = \lambda N^{[i]}
|\phi^{[i]}\rangle.
\end{equation}
Here the effective single-site Hamiltonian ${\cal H}_{\rm eff}^{[i]}\equiv (\bigotimes_{j\ne
i}^{m}\langle\phi^{[j])}|)|\Psi\rangle\langle \Psi|(\bigotimes_{j\ne
i}^{m}|\phi^{[j]}\rangle)$ is proportional to a local projector
$|\omega^{[i]}\rangle\langle\omega^{[i]}|$ at site $i$,
and the normalization $N^{[i]}\equiv\bigotimes_{j\ne
i}^{m}\langle\phi^{[j]}|\phi^{[j]}\rangle$
 is unity if all the local states are properly normalized, as will be done in practice. From the
viewpoint of, e.g., a variational Matrix Product State (MPS), one fixes all local states
$|\phi^{[j]}\rangle$ but $|\phi^{[i]}\rangle$ and solves for the
corresponding optimal $|\phi^{[i]}\rangle$ and repeats the same
procedure for $i+1$, $i+2$, etc. until the $m$-th site and sweeps
the procedure back and forth until the eigenvalue $\lambda$
converges. The converged value $|\lambda|^2$ is the square of the
maximal overlap $\Lambda_{\max}^2$. To avoid getting trapped in possible local maxima, it is useful to repeat the procedure a few times with different initial random product states and use the largest overlap obtained. 
}
\subsection{Topological Models} 

Let us now briefly revisit some of the basic properties of the models to be studied in this paper, namely the toric code, doubled semion, and doubled Fibonacci models. 

\subsubsection{Toric Code model}

The toric code  \cite{Kitaev:2003} is the simplest example of a topologically non-trivial 2d system. It is the RG fixed point of the topological phase of a
$\mathbb{Z}_2$ gauge theory, and is equivalent under local transformations to a Levin-Wen model on a honeycomb lattice
\cite{Levin:2004p117}. The Hamiltonian of the toric code is given by
\beq
H_{{\rm TC}} = -\sum_s A_s -\sum_p B_p, 
\eeq
where star operators $A_s$ and plaquette operators $B_p$ are defined as
\beq
A_s \equiv \prod_{j \in s} \sigma_x^{[j]} \ \ \ \ \ \ \ \ \ \ \ \ \  B_p \equiv \prod_{j \in p} \sigma_z^{[j]},
\eeq
with $\sigma_{\alpha}^{[j]}$ the $\alpha$-th Pauli matrix at link $j$ of the lattice. The properties of this model are well-known, including its MES \cite{Zhang:2012jc}, and its robustness to perturbations \cite{robust}. 

\subsubsection{Doubled Semion Model}

The doubled semion model \cite{doublesemion} is given by the spin model on the
honeycomb lattice
\begin{equation}
H_{\rm DS}=-\sum_s A_s -\sum_p B'_p,
\end{equation}
where $A_s$ and $B'_p$ are mutually commuting and given by
\beq
A_s \equiv \prod_{j \in s} \sigma_x^{[j]} \ \ \ \ \ \ \ \  B'_p \equiv - \prod_{k \in {\rm legs  \ of}\ p} i^{(1-\sigma_x^{[k]})/2} \prod_{j \in p} \sigma_z^{[j]}.
\eeq
As in the toric code, star and plaquette operators satisfy the non-local constraint
\beq
\prod_s A_s = \prod_p B'_p = \mathbb{I}.
\eeq
This model is known to be topologically ordered, corresponding to the universality class of a $U(1) \times U(1)$ Chern-Simons theory. The model is exactly solvable, and MES are also well known \cite{Levin:2004p117}.

\subsubsection{Doubled Fibonacci Model}

The doubled Fibonacci model \cite{Levin:2004p117} is defined on a honeycomb lattice with anyonic degrees of freedom on its edges. These degrees of freedom have two different states (say, $\ket{0}$ and $\ket{1}$), i.e., like a spin-1/2 model, but the overall Hilbert space is restricted to configurations that satisfy the Fibonacci branching rules at every vertex: 
\beqa
0 \times q = q \times 0 &=& q ~~{\rm for}~~q \in \{ 0, 1 \} , \\ \nonumber 
1 \times 1 &=& 0+1.
\eeqa
The Hamiltonian of the model is given by
\beq
H_{{\rm DF}} = -\sum_p \delta_{\Phi(p),0}, 
\eeq
where $\delta_{\Phi(p),0}$ is a projector on the states with zero flux $\Phi(p)$ through plaquette $p$. Again, the model is exactly solvable and many of its properties are well-known. 

\section{MES from GE: small sizes}

\begin{figure}
 \includegraphics[width=0.25\textwidth]{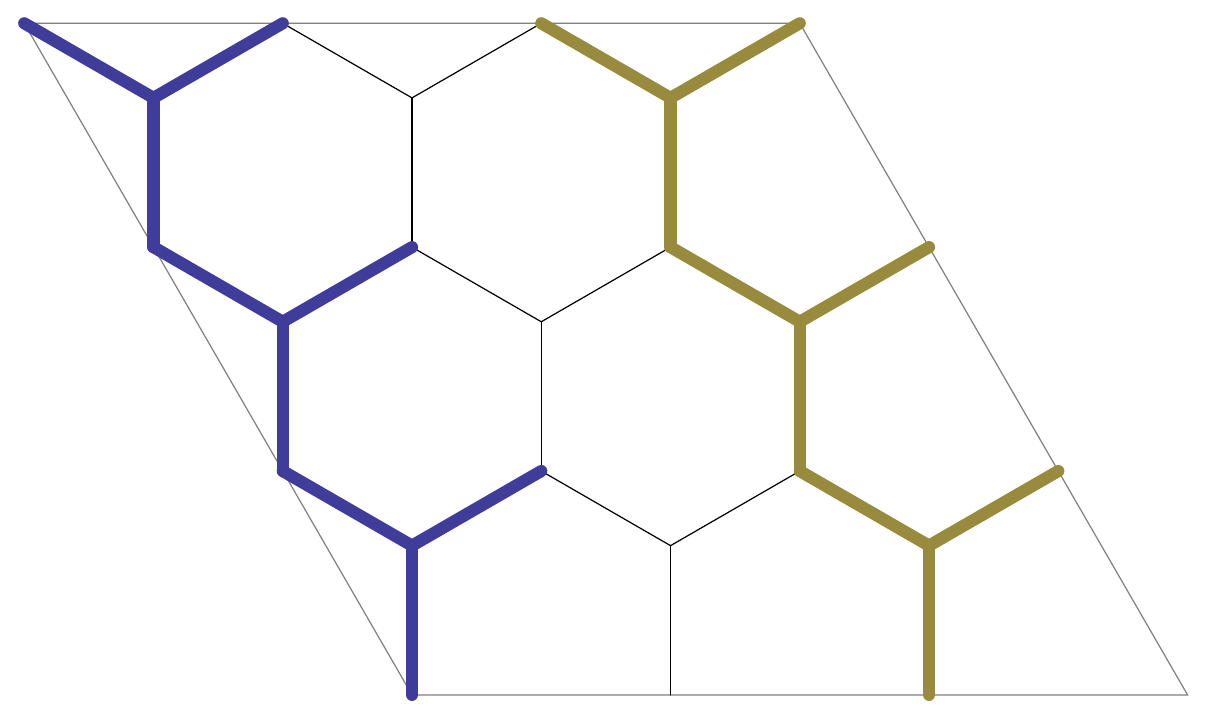}
    \caption{(Color online) $3 \times 3$ honeycomb lattice on a torus, where the unit cell is composed of the three spins around a vertex, and projected on the 2d plane. On the whole, there are 27 spins. The lattice is partitioned in 3 vertical cylinders. Spins on each cylinder have a different color on the link (blue, black, golden).}
  \label{FigLatTopo}
\end{figure}

In this section we consider the doubled semion and doubled Fibonacci models \cite{Levin:2004p117}. Both models have a topological ground state degeneracy of $4$ when placed on the torus, hence $\setbuilder{\ket{\Xi_i}} {0\leq i\leq 3}$. Small-size calculations were also performed for the toric code model \cite{Kitaev:2003}, but these produced equivalent results to those of the doubled semion model and are therefore not reproduced here. We shall, however, come back to the toric code model when considering large lattice sizes with tensor network methods in Sec.IV.  For the geometric entanglement we partition the torus into $n_\mathrm{h}$ ($n_\mathrm{v}$) cylinders of equal width in horizontal (vertical) direction and define product states with respect to those cylinders. Clearly, the results for $n_{\mathrm{h},\mathrm{v}}\geq 3$ cylinders are the most interesting ones since these are beyond bipartite entanglement (recall that for the case of $2$ cylinders, the GE corresponds exactly to the infinite-R\'enyi entropy, or single-copy entanglement).  Considering the three spins of a vertex as unit cell, a $3 \times 3$ honeycomb lattice on a torus would look like the one in Fig.~\ref{FigLatTopo}, i.e., with 27 spins on the whole. The lattice in this figure is also partitioned in three cylinders along the horizontal direction, for the sake of clarity. 

\subsection{Random sampling} 
{
Our first procedure for these small-size systems works as follows. Given a complete set of
four linearly independent states $\{|G_i\rangle\}$ in the ground space ($i=0,1,2,3$), we shall sample a large set of random quantum states  from this subspace
\begin{equation}
 \ket{\psi} := \sum_{j =0}^3  c_ {j}  \ket{G_j},
\end{equation}
and investigate their entanglement distribution.  Let us remark that the coefficients $c_j$ are obtained from a column of random unitary matrices in $U(4)$ sampled according to the Haar measure. If $\ket{G_j}$'s are not orthonormal, we will first orthonormalize them and then superpose them with the random coefficients $c_j$'s.  

We shall then attempt to identify the set of MES from those ground state samples with smallest GE. For the toric code and the doubled semion models, the set $\{\ket{\Xi_i}\}$ is fairly straightforward to write down, so we will actually take  $\ket{G_j}=\ket{\Xi_i}$ for convenience and thus we have the benefit of being able to directly check the accuracy of our numerical results.
}
%  $\ket{\psi_i}$'s  $\ket{\psi_i} \in {\mathbb C}^4$ (also called logical states), and encode them in the ground state space of our model on an $n\times L$ torus via
%\begin{equation}
%    \label{eq:encoding}
%    \ket{\psi_i(n, L)} \coloneqq  \sum_{j =0}^3   \lambda_{i j}  \ket{\Xi_j},
%\end{equation}
%where we exploit that the distinguished basis states~$\ket{\Xi_i}$ are mutually orthogonal.
%We then pretend we forgot the states~$\ket{\Xi_i}$, and attempt to reconstruct them from a uniformly random sample of ground states~$\set{\ket{\psi_i(n,L)}}$ via their  geometric entanglement. This is the scenario of unbiased reconstruction except that we have the side benefit of being able to quickly compare with our desired result.

\subsection{Systematic minimisation} 
{
Our second procedure consists of systematically minimising the GE in order to find the distinguished topological basis for this model.  Here we do not assume the four linearly independent ground states $\ket{G_i}$ to be  orthonormal, since orthonormality can be taken into account by considering
the overlap matrix  $C_{ij}\equiv \langle G_j|G_i\rangle$ and diagonalizing it as  $C_{ij} = (U^\dagger
\lambda U)_{ij}$, where $\lambda$ is a (positive) diagonal matrix
and $U$ is the unitary matrix that diagonalizes $C^T$.

%  This works as follows: let us suppose that we are given four
%ground states $|G_i\rangle$ ($i=0,1,2,3$), not
%necessarily orthonormal, but known to be linearly independent.
%Let us assume $C_{ij}\equiv \langle G_j|G_i\rangle= (U^\dagger
%\lambda U)_{ij}$, where $\lambda$ is a (positive) diagonal matrix
%and $U$ is the unitary matrix that diagonalizes $C^T$ (using
%transpose). 

We would like to find the set of MES~\cite{Zhang:2012jc,Zhu:2013wt,Zhu:2014jf}  characterized not by the entanglement entropy but by the GE.
First we need to find a state with minimum GE within the full four-dimensional ground space, with an arbitrary  state in it prametrized by $\sum_i
a_i |G_i\rangle$.  Finding the minimum is generally a nonlinear optimization problem and for our purpose here we employ the Nelder-Mead simplex method \cite{NelderMead}.  
Suppose that we found  a global minimum GE state $|\Psi_0\rangle=\sum_i a_i^{[0]} |G_i\rangle$, how would we  proceed for the  remaining MES?

%find a state with minimum entanglement within the full four dimensional subspace $\{\sum_i
%a_i |G_i\rangle\}$. \cite{Zhang:2012jc,Zhu:2013wt,Zhu:2014jf}.
%Suppose we found $|\Psi_0\rangle=\sum_i a_i^{[0]} |G_i\rangle$. Then
%we search in the subspace orthogonal to $|\Psi_0\rangle$ the next
%state with minimum entanglement. 

We need to impose orthogonality in the subsequent minimisation and it can be done  as follows. Suppose we have two states
$|\Psi\rangle=\sum_i a_i |G_i\rangle$ and $|\Psi'\rangle=\sum_i
b_i|G_i\rangle$. Their overlap is $\langle \Psi'| \Psi \rangle= \sum_{ij}
b_j^* a_i \langle G_j|G_i\rangle=\vec{b}^{*}\cdot C \cdot \vec{a}=
\vec{b}^* U^\dagger \sqrt{\lambda}\sqrt{\lambda} U\vec{a}$. It is
convenient to define a matrix $T\equiv \sqrt{\lambda} U$ so that
$\langle \Psi'| \Psi \rangle=\vec{b}^* T^\dagger   T\vec{a}$. Orthogonality is then imposed by restricting the parameters of $\vec{b}$ (used in the optimization program)
to those satisfying $\vec{b}^*T^\dagger   T\vec{a}=0$.

We denote by $|\Psi_1\rangle=\sum_i a_i^{[1]} |G_i\rangle$ the state with minimum entanglement in
the subspace that is orthogonal to $|\Psi_0\rangle$, using the procedure described above.  We continue to
find the  next state $|\Psi_2\rangle$  with minimum entanglement restricted to be 
orthogonal to the subspace spanned by $|\Psi_0\rangle$ and
$|\Psi_1\rangle$. This can be done via restricting 
the parameters $\vec{c}$ used in the optimization to those satisfying both $\vec{c}^*T^\dagger   T\vec{a}^{[0]}=0$ and $\vec{c}^*T^\dagger   T\vec{a}^{[1]}=0$.

}
Having found the first three orthogonal states, the
fourth state $|\Psi_3\rangle$ is automatically determined. These four states would then constitute our candidates for the basis of MES.

\subsection{Results}

\subsubsection{Doubled semion model}
{
Calculations have been performed for the \emph{same} uniformly random sample~$\mathfrak{S}$ of 196608 states $\sum_{j=0}^3 c_{ij} \ket{\Xi_j}$ from the same set of random unit vectors in $\mathbb{C}^4$ but for different system sizes. These states are called logical states because they are effectively encoded two-qubit states on a torus. For 3 or more cylinders we obtain an approximation in terms of numerical upper \emph{bounds} on the GE via 10 lower bounds on the overlap~\footnote{The reason we refer to these as lower bounds is because numerically we cannot certify that the overlap from our procedure is absolutely maximum, hence only a lower bound on GE.}.
}
 In order to obtain each lower bound, we draw a random product state and update the overlap one cylinder (party) at a time until convergence is reached. This algorithm yields a non-decreasing sequence of overlaps, i.e. a local maximum of the overlap, and is in fact the usual approach used also in tensor network methods to compute approximations to the GE~\cite{topoGE2}. 

\begin{figure}
 \includegraphics[width=0.5\textwidth]{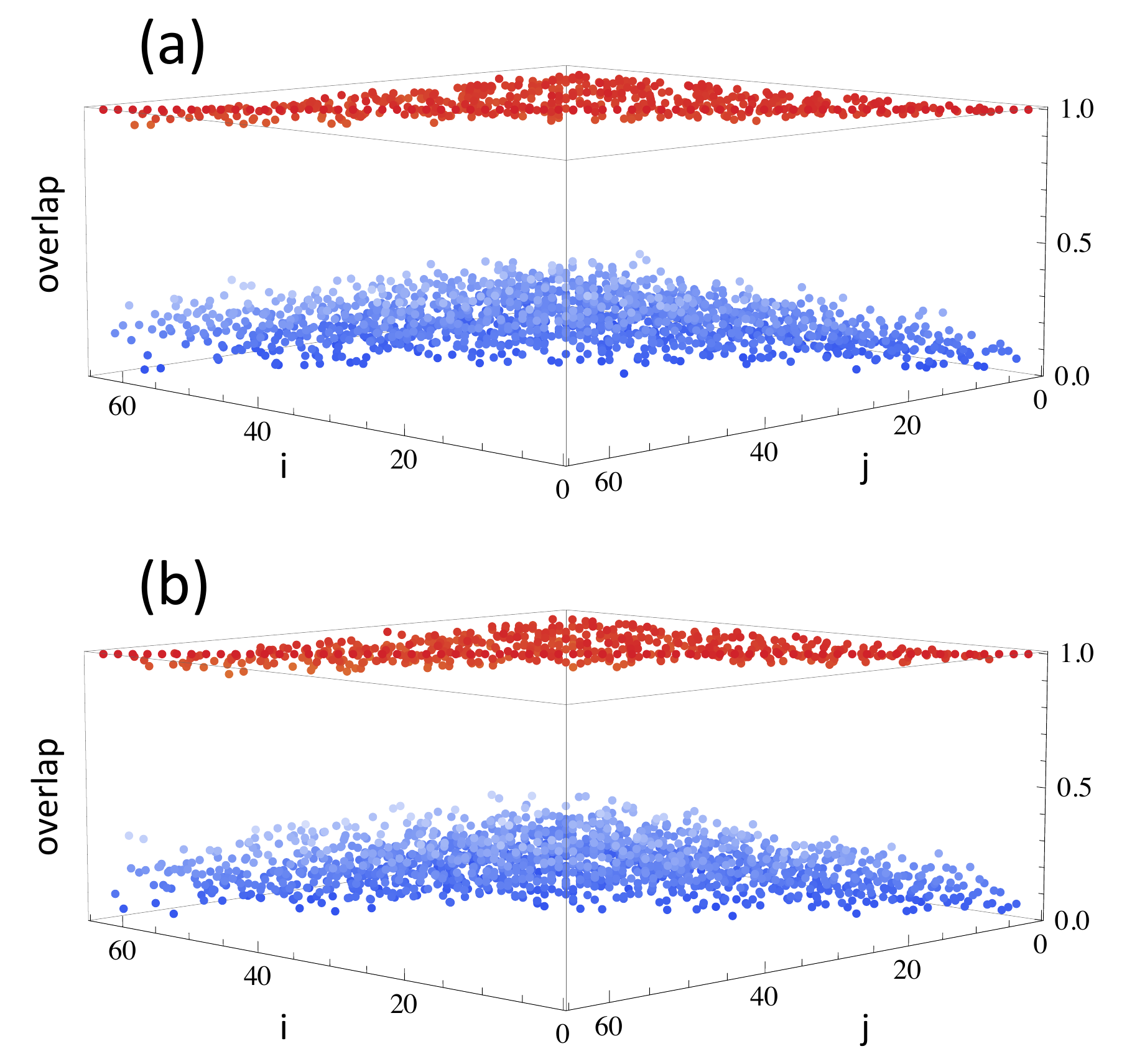}
    \caption{(Color online) The overlaps $\abs{\braket{\psi_i}{\psi_j}}$ (for $i\geq j$) of the 64 logical states with smallest \textsc{GE} for the doubled semion model, for (a) horizontal and (b) vertical cylinders on $2\times 2$, $3\times 3$ and $3\times 4$ honeycomb lattices (27 and 36 spins, respectively). The logical states with lower GE do not depend on the size of the lattice, hence we get the same two plots for the three sizes.}
  \label{fig:overlaps}
\end{figure}

\bigskip

\underline{\emph{(a) Small GE states.-}}

\bigskip

For each encoded state of the sample~$\mathfrak{S}$, we compute its total GE for a partition into cylinders for some given size of the system, and then sort the states by increasing GE value. Importantly, we have seen that the permutation to perform this sorting, for a given direction of the cylinders, is independent of the system size. Once the states are sorted, we pick those states with smallest GE, and study their properties. 

First we analyze how orthogonal the small \textsc{GE} states are. Since the $\ket{\Xi_j}$ are orthonormal the inner product evaluates to $\braket{\psi_i}{\psi_j} = \sum_{k=0}^3 c_{ik}^* c_{jk}$.  The overlaps are shown in Fig.~\ref{fig:overlaps}. Clearly these are either very close to one, or very close to zero. This picture is consistent with low-entanglement states: whenever the overlap of two states is close to one we interpret that these states are close to the same MES $\ket{\Xi_i}$, whereas whenever the overlap is close to zero we interpret that these are close to two different and orthogonal MES $\ket{\Xi_i}$ and $\ket{\Xi_j}$. 

Given this we can extract a quasi-orthogonal basis for the ground state subspace from the smallest \textsc{GE} states. We do this by separating the states into sets of $k$ states which are almost identical, i.e., the overlap of any two states within a set is $ \ge 0.9$, and choosing the smallest $k$ that yields four sets. Then, from each set we pick the state with the smallest \textsc{GE}. In this way we build our four candidates for MES. We find that these always have almost maximum overlap with exactly one of the four states $\ket{\Xi_i}$, and almost zero with the remaining ones. In this way, we certify that we indeed found a very good approximation to the correct basis of MES by looking at states with minimal \textsc{GE}. 

{
In fact, the basis of MES found in this way agrees, with good accuracy, with the one found using the systematic minimisation procedure described previously (and will be discussed in more detail for the case of the doubled Fibonacci model). Using systematic minimisation, we report here that we have successfully identified the correct four MESs, all with $E_G=6$ on the $3\times3$ honeycomb (same for the toric code), partitioned into three cylinders. From this set of states we can extract the modular matrices $S$ and $T$, if needed.  It is important to stress that the four MES are the only four with the lowest GE values; any other states than the four have higher values of GE.
}
\begin{figure}
 \includegraphics[width=0.45\textwidth]{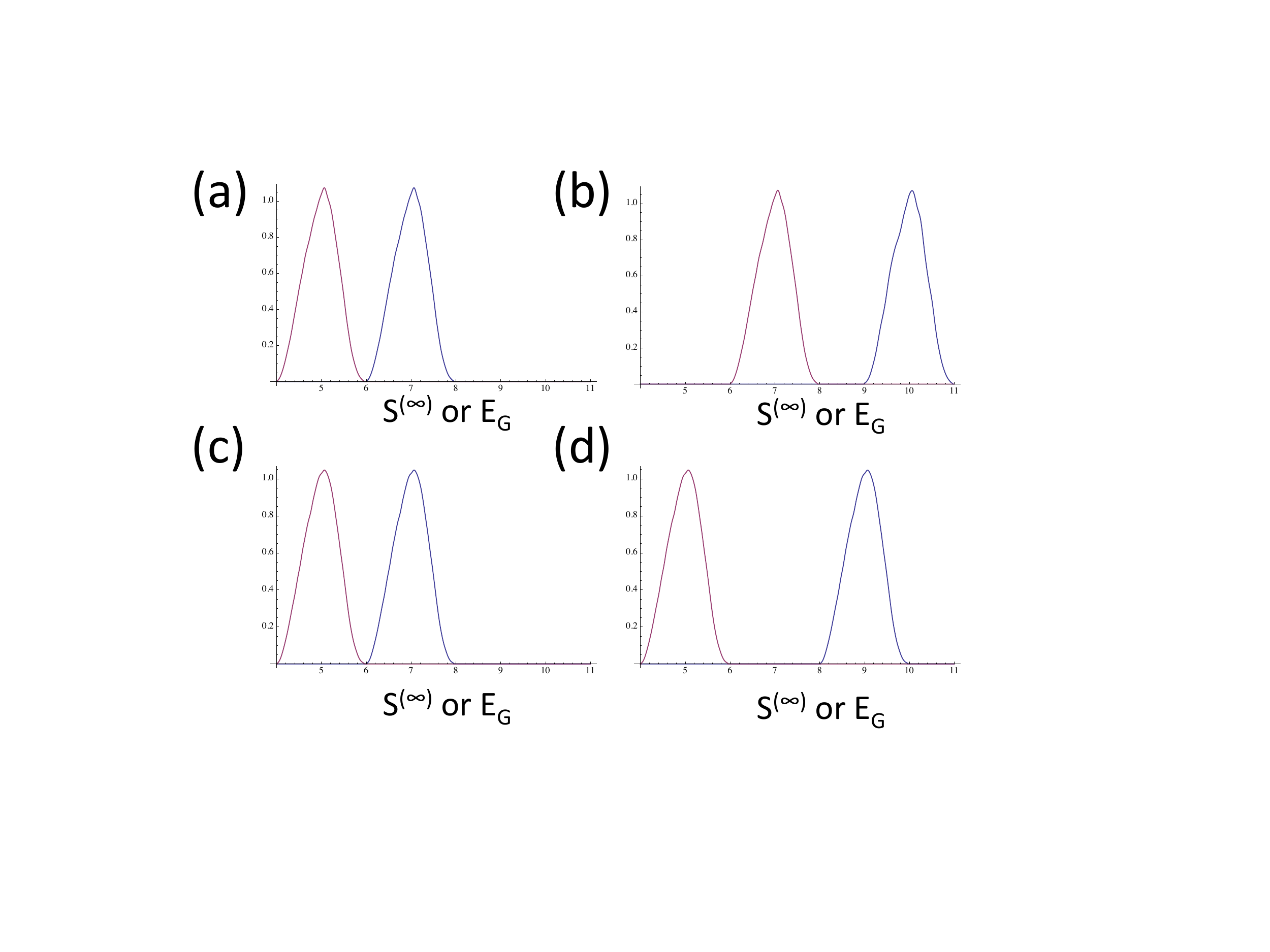}
    \caption{(Color online) The \textsc{GE} (blue) versus single-copy entanglement (red) histograms for horizontal and vertical cylinders on honeycomb lattices of different sizes, for the doubled semion model: (a) 27 spins and 3 vertical cylinders, (b) 36 spins and 3 vertical cylinders,  (c) 27 spins and 3 horizontal cylinders, and (d) 36 spins (as in (b))  and 4 horizontal cylinders.}
        \label{fig:histograms_minentropy}
\end{figure}

We conclude, thus, that the \textsc{GE} correctly identifies the quantum states of the topologically distinguished basis. Let us stress that similar results were also found for the toric code model (not shown). 

\bigskip

\underline{\emph{(b) Comparison to single-copy entanglement.-}}

\bigskip

\label{sec:ge_min_entropy}

One could be tempted to say that the \textsc{GE} of a state with respect to a partition into, say, $n$ cylinders on a torus, should be approximately given by $E_G \sim 2(n-1)S^{(\infty)}$ where $S^{(\infty)} = - \log \nu_1$ is the infinite-R\'enyi entropy or single-copy entanglement \cite{sc}, and $\nu_1$ is the largest eigenvalue of the reduced density matrix of a bipartition into two cylinders. This is inspired by a picture based on matrix product states (\textsc{MPS}) \cite{tn}, where the two boundaries of each cylinder should effectively decouple \emph{up to a topological contribution}, once the cylinder is wider than the correlation length. Clearly, for $n=2$ cylinders $S^{(\infty)}$ is the exact GE value, so we should regard the GE as a natural multipartite generalisation of the bipartite infinite-R\'enyi entropy. Still, the relation between them in the multipartite scenario \emph{and} for topological models is not so clear. 

Here we try to throw and cast a bit of light on this question, by comparing the values of $E_G$ and $S^{(\infty)}$ for the states in our sample, for small lattice sizes of the doubled semion model. In Fig.~\ref{fig:histograms_minentropy} we can see a comparison of the histograms of states of our sample, i.e., the ratio of states with a given value of either the GE or the single-copy entanglement, for different partitions and lattice sizes, and cylinders of one lattice-site width. Our results seem to be in agreement with a global offset between the GE and the single-copy entanglement for all the states in the sample (which correspond to different ground states), far away from the conjectured (quasi)-decoupled regime for wide cylinders. In fact, our data for this model is consistent with the behaviour 
\beq
        E_G   = S^{(\infty)}+ (n- 2) (L-1)  ~~~{\rm (conjecture)}
\eeq
for cylinders of one unit cell width, where $n$ is the number of cylinders and $L$ their circumference (in unit cells). Such a dependence might very well be different once wider cylinders are considered, once a perturbation is added to the topological model, and once the spectrum of eigenvalues of the reduced density matrix stops being flat. It would also be of interest to know if such a relation, or similar, holds for models with more complex types of topological order. 

{
Finally, let us remark that we do not necessarily need to restrict ourselves to consider the torus to be obtained from a square, but also $n$ cylinders with each having length $L$ and width $w$.
}

\subsubsection{Doubled Fibonacci model}

In order to analyze the properties of this model we used similar methods as in the previous section. However, and as we shall see, the systematic minimisation method becomes quite important here since some of the states in the topologically distinguished basis do not correspond to a global minimum in entanglement, given the non-Abelian nature of the system.  
{This is in contrast to what we have found earlier in the toric code and doubled semion models, where all the states in the topologically distinguished basis, i.e., the MES, have the same global minimum value of GE.
}

\begin{figure}
       \includegraphics[width=0.45\textwidth]{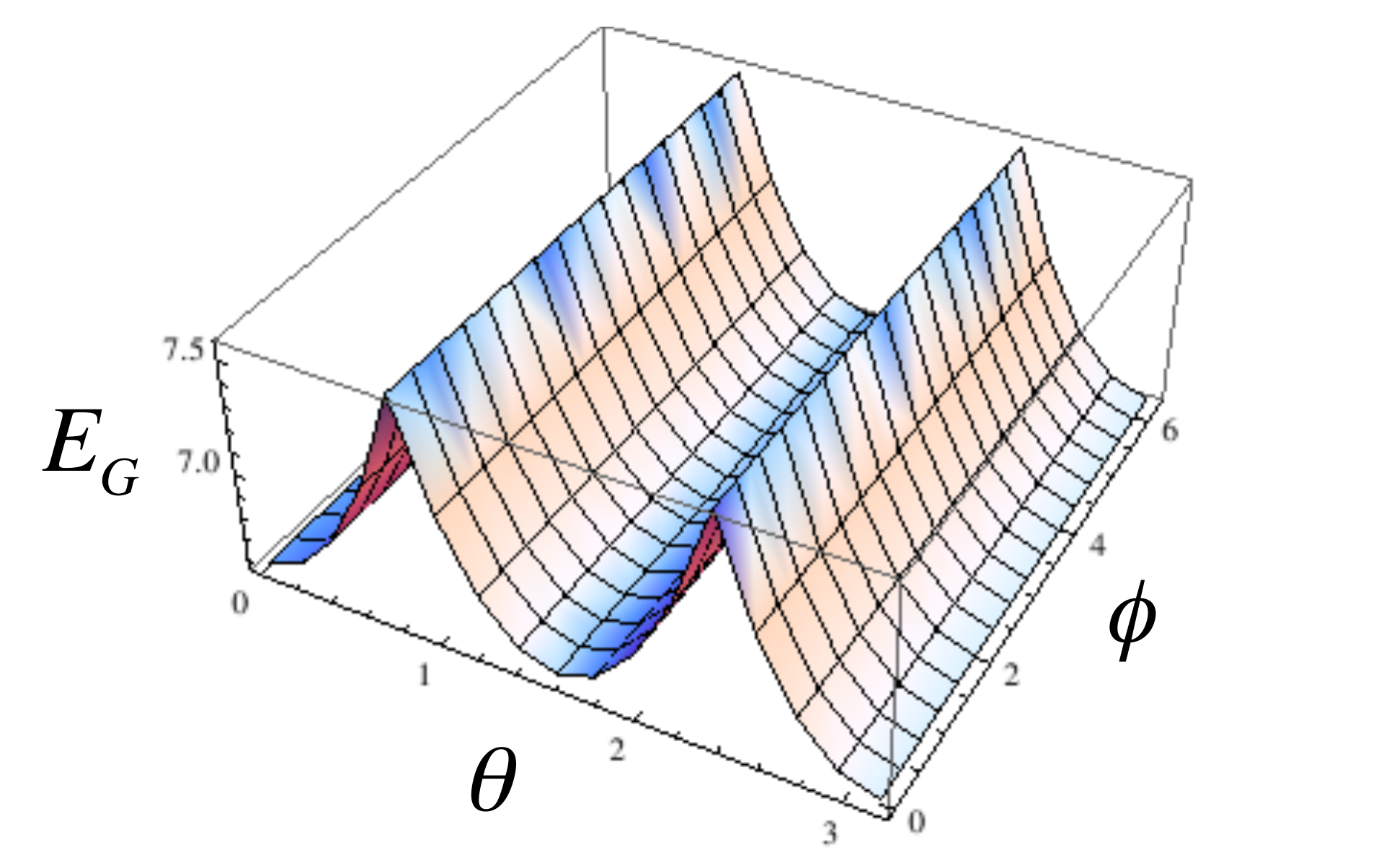}
      \caption{(Color online) GE in the linear superposition of $\cos\theta|\Psi_1\rangle+\sin\theta\,e^{i\phi}|\Psi_2\rangle$ for 3 cylinders. 
      At $\theta=0,\pi/2$, the corresponding states are $\ket{\Psi_1}$ and $\ket{\Psi_2}$, respectively (regardless of $\phi$). The figure shows that these two states are the only two minimum entangled states in this subspace, and any other superposition will necessarily have higher entanglement. Curiously, the entanglement is independent of the angle $\phi$.}   
         \label{fig:states23H2x2}
\end{figure}

\bigskip

\underline{\emph{(a) Small GE states and non-Abelian character.-}}

\bigskip

{
Earlier, we have applied the first method of random sampling to the doubled Fibonacci model and found that, unlike in the toric code and doubled semion models,  there is only \emph{one} global minimum \textsc{GE} state. Upon closer inspection this fact is less surprising, since the corresponding entanglement \emph{entropy} depends on the quantum dimension of each distinguished basis state $\ket{\Xi_i}$. It is known that for this model the ground-state space $\mathcal{L}$ decomposes as $\mathcal{L}_1\oplus\mathcal{L}_\phi\oplus\mathcal{L}_{\phi^2}$ (where $\phi$ is the golden section), where the dimension of $\mathcal{L}_1$ and $\mathcal{L}_{\phi^2}$ is one and the dimension of $\mathcal{L}_\phi$ is two. From such a structure we may expect to find three distinct GE values for the MES, i.e., with the two topologically distinguished states in $\mathcal{L}_\phi$ having the same value of GE. If our observation is true, then  this implies that the \textsc{GE} can actually be used to extract the number of  quasiparticle excitations in each sector from the number of quasi-orthogonal states (i.e. subset of MES) and their  \textsc{GE} values.  

Motivated by the previous observation, we thus carry out the systematic minimisation procedure in order to find the distinguished topological basis for this model. Specifically, we considered lattices of 27 ``spins'' partitioned into 3 cylinders. Proceeding in this way, the four MES obtained via GE have
very similar (within less than 1\% error) coefficients to those
obtained via the entanglement entropy in a $2\times2$ system. 
The specific values of GE that we find for these four states are $E_G(\Psi_0) = 4.8443$, $E_G(\Psi_1) = 6.5698$, $E_G(\Psi_2) = 6.5698$, and $E_G(\Psi_3) = 7.7303$. The fact that $|\Psi_1\rangle$ and $|\Psi_2\rangle$ possess almost the same GE confirms our expectation that there are indeed two MES with the same GE value, implying $|\Psi_1\rangle$ and $|\Psi_2\rangle$ are in
$\mathcal{L}_\phi$. As an illustration we plot in Fig.~\ref{fig:states23H2x2}  the numerical landscape of entanglement in the $\mathcal{L}_\phi$ subspace, showing interesting features.
Moreover,
it appears that in our basis $|\Psi_2\rangle=
(|\Psi_1\rangle)^*$. Moreover,  $|\Psi_0\rangle
\in\mathcal{L}_1$ with the lowest entanglement, and $|\Psi_3\rangle \in \mathcal{L}_{\phi^2}$ having the largest entanglement among the MES.

Furthermore we find the product of modular matrices $TS$ to be accurate up to an error of $< 1\%$, once global phases of the MES have been fixed:
\begin{widetext}
\beq
TS \approx 
\begin{pmatrix}
+0.276 + 0.003 i &+ 0.448 &   +0.445 + 0.001 i & + 0.724 - 0.001 i \\
-0.362 - 0.264 i &   +0.224 + 0.160 i &-0.585 - 0.427 i & +0.361 + 0.262 i\\
 -0.363 + 0.261 i & -0.586 +    0.424 i & +0.224 - 0.163 i & + 0.362 - 0.263 i \\
 +0.723 - 0.002 i & -0.448 + 0.003 i & -0.447 + 0.001 i & +0.276 
\end{pmatrix}.
\eeq
\end{widetext}
}

\section{MES from GE: large sizes}

To extend the study of MES to larger system sizes, we employ 
a tensor network construction of the ground state. In our case we use PEPS, 
which have proved to represent faithfully ground states of 
topological models \cite{topoPEPS}. In some situations, an analytic derivation of
the tensors can be obtained directly from the Hamiltonian. In general, though, the tensors 
are obtained after a numerical optimisation of the energy. The numerical contraction of 
a PEPS allows a precise and efficient calculation of the 
geometric entanglement even in the case of blocks, as explained in Ref.~\cite{topoGE2}. 

\subsection{Reminder of the tensor network method}

\begin{figure}
\includegraphics[width=0.5\textwidth]{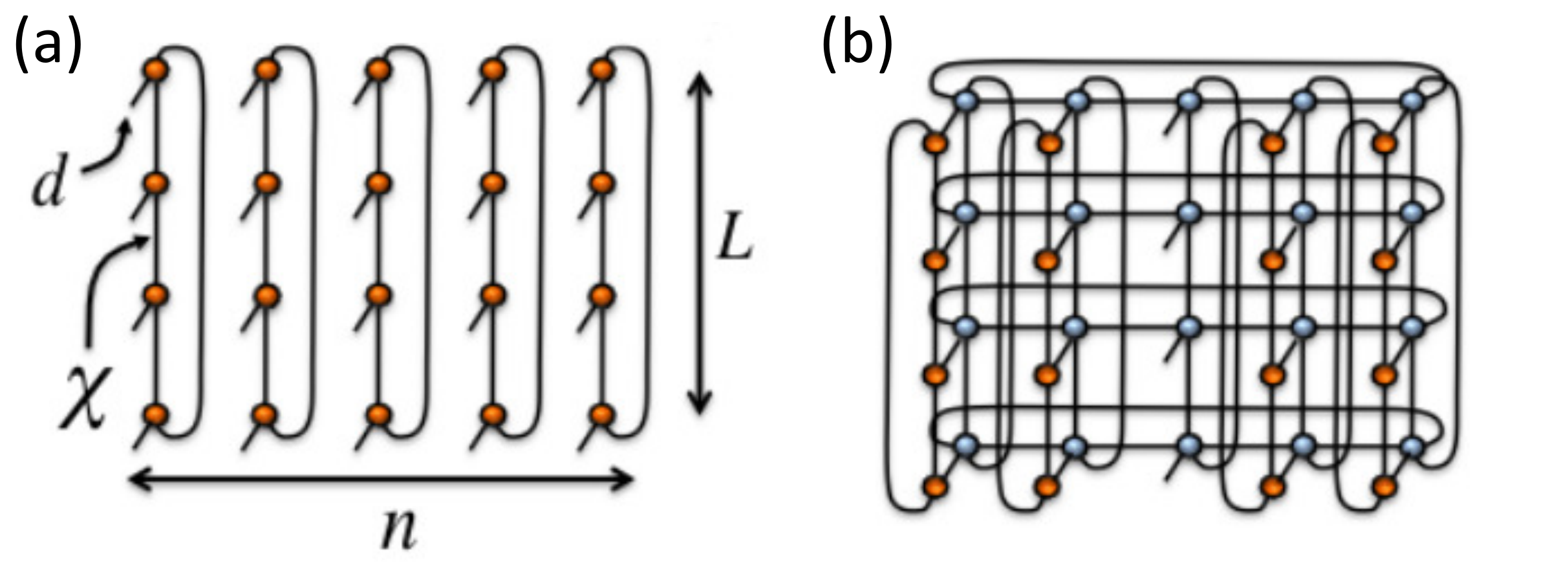}
\caption{(Color online) Diagrammatic representation of the optimisation performed to obtain the GE:
(a) we first set up a random set of 1d MPS $|\phi^{[i]}\rangle$ of bond dimension $\chi$ for all cylinders $i$ forming $|\Phi\rangle$. Each 
of these states covers the torus along one direction. (b) At each step we pick a position $k$ 
where we obtain a new $|\phi^{[k]}\rangle$ after contracting the rest of the torus tensors. This 
procedure is repeated sweeping back and forth along the torus until convergence is obtained. In the figure, $d$ is the physical dimension of the sites.} 
\label{fig:peps_diag}
\end{figure}

We set up a PEPS on a square lattice with periodic boundary 
conditions. The tensors represent the state $|\Psi(n,L)\rangle$, 
where $n$ and $L$ are the sizes of the torus defined by the boundary conditions. 
The GE is obtained from the set $|\Phi\rangle$ of $n$ states 
of length $L$ that cover the entire torus across cylinders. These states are one-dimensional  
and have periodic boundary conditions, therefore we approximate them by MPS with periodic boundary conditions, see Fig.~\ref{fig:peps_diag}(a). 

Our goal is the optimisation of the MPS states so that the overlap
$$
\Lambda_{max} = \frac{|\langle\Phi|\Psi\rangle|}{\sqrt{\langle\Psi|\Psi\rangle\langle\Phi|\Phi\rangle}}
$$
is maximised. The quantities $\langle\Phi|\Psi\rangle$ and $\langle\Psi|\Psi\rangle$ can be efficiently estimated 
using standard PEPS methods. For large systems (say, $L>20$) this calculation can be achieved very efficiently using the MPS description for each of the states $|\phi^{[i]}\rangle$ \cite{topoGE2}. 

We remind here how the optimisation of the product state $|\Phi\rangle$ 
can be performed for a torus of size $n\times L$ (see Fig.~\ref{fig:peps_diag}(b) as a reference).
The key observation is that the optimisation is performed on each $|\phi^{[i]}\rangle$ iteratively,
optimising a single state in each step and sweeping along the torus until convergence is reached.
Following Ref.~\cite{topoGE2}, the procedure can be summarised as follows:

\bigskip

\emph{1.-} We start with a random choice for each $|\phi^{[i]}\rangle$ as the initial state of the optimisation.

\bigskip

\emph{2.-} Choose a position $k$, and by fixing all the remaining states optimise a new $|\phi^{[k]}\rangle$.
This state is obtained after contracting the entire PEPS, keeping the tensor structure to form the new $|\phi^{[k]}\rangle$.
For small systems one can perform a single contraction of the tensors for each position $i \neq k$ and continue the optimisation
as a purely 1d problem.

\bigskip

\emph{3.-} Move to position $k+1$. Sweep along the torus iterating these steps until convergence in the overlap is obtained.
In our simulations convergence is obtained after a few sweeps along the torus.

\bigskip

\subsection{Results: toric code model}

As an archetypical example of topological order we analysed here the toric code model with our tensor network method. Since this is the simplest 2d model with topological order, we have a very good control over its representation in terms of PEPS. We focus our attention on the toric code model 
on a square lattice with periodic boundary conditions. 

\bigskip

\underline{\emph{(a) Small GE states from tensor networks.-}}

\bigskip

The contraction of PEPS allows a precise calculation of 
the GE to be used in order to identify the 
MESs. Using the tensor network construction, the states $|00\rangle$
and $|10\rangle$ (we follow the notation introduced in Ref.~\cite{topoGE1}) 
can be created easily in the PEPS picture with a small bond dimension $D=2$.
The MES $|\Xi_0\rangle =2^{-1/2}(|00\rangle+|10\rangle)$ can easily be constructed using a 
bond dimension $D=4$. These conditions allow an optimal description of the problem as a PEPS,
and result in an efficient iterative search of the maximal overlap.

\begin{figure}
\includegraphics[width=0.3\textwidth]{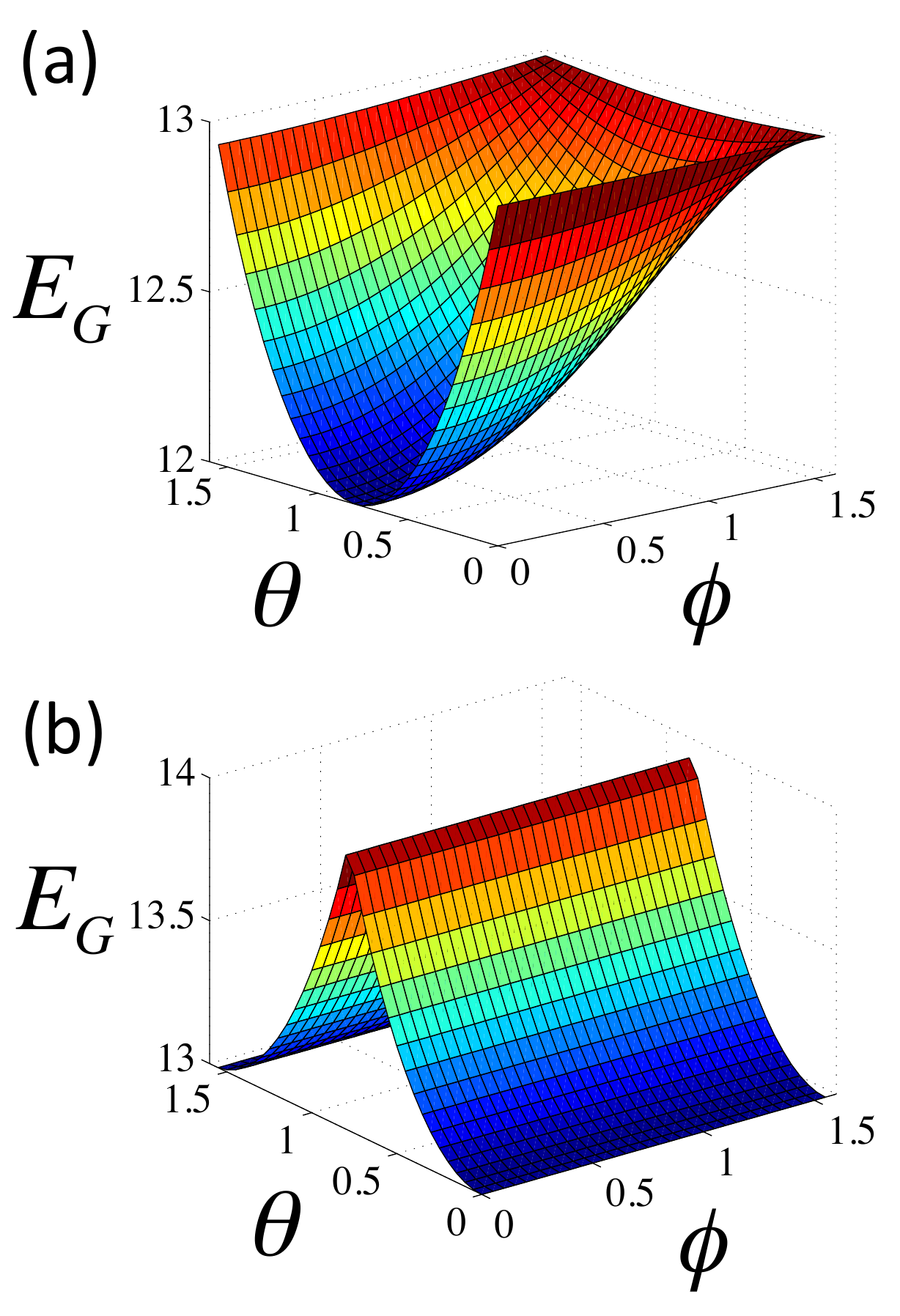}
\caption{(Color online) (a) GE for the state $|\theta,\phi\rangle = \cos\theta|00\rangle + \sin\theta\,e^{i\phi}|10\rangle$.
We find a minimum at $\theta=\pi/4$ corresponding to $|\Xi_0\rangle =2^{-1/2}(|00\rangle+|10\rangle)$. (b) 
The superposition of states $|00\rangle$ and $|01\rangle$ yields a larger value of the GE for any value of \
$\theta$ and $\phi$, with the maximum located at precisely $\theta=\pi/4$.}
\label{fig:ge_peps}
\end{figure}

In order to show how to the GE can be used to clearly identify a MES with the tensor network approach for large system sizes, we 
study the parametrization 
$$
|\theta,\phi\rangle = \cos(\theta)|00\rangle + \sin(\theta)e^{i\phi}|10\rangle.
$$
The GE for this state is represented in Fig.~\ref{fig:ge_peps}(a) as a function of both $\theta$ and $\phi$. 
The minimum of the GE appears at $\theta=\pi/4$, corresponding to 
the superposition $|\Xi_0\rangle =2^{-1/2}(|00\rangle+|10\rangle)$.
A similar exploration can be performed for the superposition of $|00\rangle$ and $|01\rangle$, which is shown in 
Fig.~\ref{fig:ge_peps}(b). In this case however, no MES is found and the GE increases for any value of 
$\theta$ and $\phi$, as expected from entropy calculations \cite{Zhang:2012jc}. The two plots in Fig.~\ref{fig:ge_peps} are obtained for a system of $4\times 4$ sites. 
Identical results are obtained for larger systems up to some overall displacement in the amount of GE due only to the size of the system. In practice, we checked this for systems up to $\sim 20 \times 20$, with no change in the conclusions. 

\begin{figure}
\includegraphics[width=0.33\textwidth]{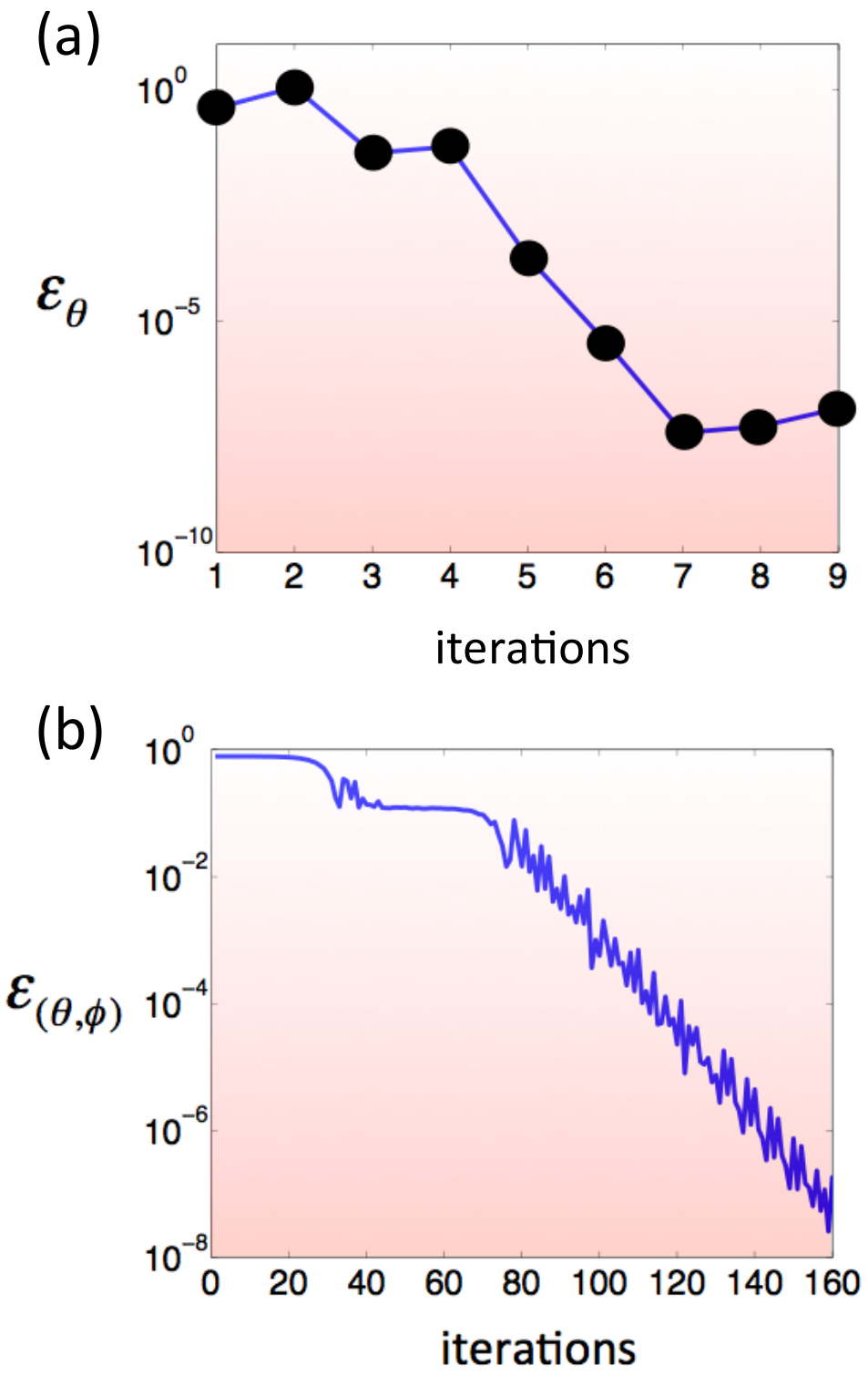}
\caption{(Color online) 
{
Evolution of the absolute error in the gradient method while searching for the state with minimum GE 
in the space defined by $\theta$ and $\phi$ in $|\theta,\phi\rangle = \cos\theta|00\rangle + \sin\theta\,e^{i\phi}|10\rangle$, with (a) $\epsilon_\theta \equiv |\theta-\pi/4|$ (fixing $\phi = 0$) and (b) $\epsilon_{(\theta, \phi)} \equiv | (\theta, \phi) - (\pi/4,0)|$. The size of the system is $4\times 20$.}}
\label{fig:ge_peps2}
\end{figure}

\bigskip

\underline{\emph{(b) Systematic minimisation with tensor networks.-}}

\bigskip

Since we have a way to determine the GE for large lattices using PEPS, we can actually identify the MESs via GE using some numerical minimisation algorithm. We checked this by optimising over the 2-parameter space spanned by $\theta$ and $\phi$ shown in Fig.~\ref{fig:ge_peps}. This parameter space 
can be explored by, e.g., a gradient method, in order to identify the minimum GE corresponding to MES.

We show in Fig.~\ref{fig:ge_peps2} the result of such an optimisation: starting from a random state, the value of GE evolves along the optimisation for the 
superposition of $|00\rangle$ and $|10\rangle$ in a system of size $4\times 20$. The optimal value $\theta_{opt}=\pi/4$ is obtained 
with an error $<10^{-8}$ after only a few iterations, being of the order of the precision imposed on the minimiser. Even though we only show here the combination 
of two states of the ground state space, our results clearly suggest that this process can be extended to the full basis of ground states and to larger parameter spaces in order to perform a full search if required. 

\section{Conclusions and outlook}

In this paper we have shown that the GE can be used as a powerful and useful tool to identify the distinguished basis of MES on a 2d system with topological order. We have seen this with  calculations for the toric code, doubled semion and doubled Fibonacci models. Large-scale calculations for the toric code model have been done using a recently proposed tensor network approach \cite{topoGE2}. Our results for the doubled Fibonacci model also show that, indeed, it is possible to read off the number of Abelian quasiparticle excitations in the topological model directly from the optimisation procedure. Moreover, the results of this paper provide a very straightforward and efficient way of determining the topological properties of a strongly correlated system, especially when combined with the tensor network numerical approach. It would be very interesting to apply the ideas and methods discussed in this paper to other interesting 2d topological models, especially those having non-Abelian quasiparticle excitations. This will be the subject of future investigations. 

%\acknowledgements

\bigskip

{\bf Acknowledgements:} Discussions with B. Bauer, F. Pollmann and G. Vidal are acknowledged. T.-C.W. and A.G.-S. acknowledge the support by the National Science Foundation under Grants No. PHY 1314748 and No. PHY 1333903. This research was supported in part by Perimeter Institute for Theoretical Physics. Research at Perimeter Institute is supported by the Government of Canada through Industry Canada and by the Province of Ontario through the Ministry of Research and Innovation.

\end{document}